\documentclass[conference]{IEEEtran}
\usepackage{cite}
\usepackage{amsmath,amssymb,amsfonts}
\usepackage{algorithmic}
\usepackage{graphicx}
\usepackage{textcomp}
\usepackage{xcolor}
\def\BibTeX{{\rm B\kern-.05em{\sc i\kern-.025em b}\kern-.08em
    T\kern-.1667em\lower.7ex\hbox{E}\kern-.125emX}}
\begin{document}

\title{Simple 1-D Convolutional Networks for Resting-State fMRI Based Classification in Autism\\
}

\author{
\IEEEauthorblockN{1\textsuperscript{st} Ahmed El Gazzar}
\IEEEauthorblockA{\textit{Department of Psychiatry} \\
\textit{AMC, University of Amsterdam}\\
Amsterdam, The Netherlands \\
gazzar033@gmail.com} 
\and
\IEEEauthorblockN{2\textsuperscript{nd} Leonardo Cerliani}
\IEEEauthorblockA{\textit{Department of Psychiatry} \\
\textit{AMC, UvA}\\
Amsterdam, The Netherlands \\
leonardo.cerliani@gmail.com}
\and
\IEEEauthorblockN{3\textsuperscript{rd} Guido van Wingen}
\IEEEauthorblockA{\textit{Department of Psychiatry} \\
\textit{AMC, UvA}\\
Amsterdam, The Netherlands \\
guidovanwingen@gmail.com}
\and
\IEEEauthorblockN{4\textsuperscript{th} Rajat Mani Thomas}
\IEEEauthorblockA{\textit{Department of Psychiatry} \\
\textit{AMC, UvA}\\
Amsterdam, The Netherlands \\
rajatthomas@gmail.com}
}

\maketitle

\begin{abstract}
Deep learning methods are increasingly being used with neuroimaging data like structural and function magnetic resonance imaging (MRI) to predict the diagnosis of neuropsychiatric and neurological disorders. For psychiatric disorders in particular, it is believed that one of the most promising modality is the resting-state functional MRI (rsfMRI), which captures the intrinsic connectivity between regions in the brain. Because rsfMRI data points are inherently high-dimensional ($\sim$1M), it is impossible to process the entire input in its raw form. In this paper, we propose a very simple transformation of the rsfMRI images that captures all of the temporal dynamics of the signal but sub-samples its spatial extent. As a result, we use a very simple 1-D convolutional network which is fast to train, requires minimal preprocessing and performs at par with the state-of-the-art on the classification of Autism spectrum disorders.  
\end{abstract}


\section{Introduction}
\label{intro}
Deep learning has, in recent years, achieved enormous success in a multitude of computer vision tasks including classification, regression, segmentation, and tracking, to name a few. Within deep learning, the convolutional architectures have proven especially effective because of their ability to leverage translational symmetry and parameter sharing. The neuroimaging community have adopted these techniques to good effect. In \cite{b1}, the authors describe the myriad of ways in which deep learning has been successfully utilized to classify psychiatric and neurological disorders using neuroimaging.\\
\\
There are a number of challenges when it comes to applying deep learning to neuroimaging studies; (i) dimensionality: each data point is either a high resolution structural MRI (sMRI) or a 4-D volume of resting-state functional MRI (rsfMRI). These data points have from 0.1 to 1 Million voxels (volume elements), (ii) small sample-size: data from any single research centre or hospital is typically in the range of 10 to about a 100 subjects. Deep learning on the other hand is a data-hungry technique often requiring thousands of labelled data points.\\
\\
In this paper, we propose a simple method to transform the high-dimensional rsfMRI data to a few time-series, which can be used to train a 1-D convolutional network for classification. This approach does not require to reduce the information of the time-series into summary measures such as temporal correlations, mean, variance, regional homogeneity, and amplitude of low frequency fluctuations. To test our approach we used the ABIDE autism database. Autism is a developmental neuropsychiatric disorder affecting $\sim$1\% of the population. The pathophenotype of autism includes a combination of deficits in social communication and stereotyped/repetitive behaviours - including abnormal reactivity to sensory stimuli - and narrowly focused interests. However, there is very high variability across patients with respect to the modality and degree of manifestation of these deficits.\\
\\
Similar to other neuropsychiatric conditions, in autism there appear to be no macroscopic differences in brain morphology. On the other hand, histological, neurophysiological and neuroimaging studies in the last 20 years strongly suggested that autism is characterized by differences in brain connectivity. Still to date very few of the reported results have been consistently replicated in a substantial number of studies. The limited sample size of traditional experiments (in the range of 20-100 subjects) likely contributed to generate results which were difficult to generalize to a large population.\\
\\
The recent release of the ABIDE autism neuroimaging dataset (details in \ref{preprocess}), featuring 2000+ structural and functional MRI scans of autistic (ASD) and typically developing (TD) participants from 29 different centers, has the potential to yield the discovery of robust neuroimaging biomarkers of autism, which can be generalized to patients within a wide phenotypical spectrum and to different acquisition centers and scan procedures.\\
\\
In this paper we first describe the prior work done on the classification of the ABIDE data set using machine learning in \ref{pwork}. In \ref{preprocess} we detail the data set and the preprocessing pipeline used. In section \ref{app} we outline the approach to creating the time-series for regions-of-interest (ROI) defined using different brain atlases. The experiments performed and the results are in \ref{expnres} and finally in \ref{con} we discuss the findings and suggest the steps forward.

\section{Prior work and our contribution}
\label{pwork}
Previous works which have tried to predict the diagnosis of autism using traditional machine-learning algorithms (or deep neural networks) have focused on the similarity (or correlation) matrix of the rsfMRI signal, averaged over different brain locations using an atlas in standard space. The feature vector of each subject is generally constructed by vectorizing the upper triangular part of the similarity matrix. Using this approach, \cite{b14} used support vector classification (SVC) and ridge regression on the ABIDE-I (N=1112, ref. \ref{preprocess} for details of the ABIDE) and compared the effect of different atlases (10-200 regions) and of different similarity measures (linear correlation, partial correlation and tangent embedding) to achieve a final intra-site classification accuracy of 67\%. In \cite{b17} the authors used a multi-linear perceptron (MLP) preceded by two stacked denoising autoencoders to achieve a final across-sites accuracy of 70\%, outperforming support vector machines (SVM) (65\%) and radom-forests (RF) (63\%). \\
\\
A few other studies using deep learning employed graph convolution approaches (GCN), where each node of the the graph represent a participant, and the signal on each of node is associated with a neuroimaging-derived feature vector, while the edges of the nodes incorporate phenotypic information. In this context, \cite{b15, b20} achieved a maximum accuracy of 70.4\% and 70.86\%, using as feature vector(s) either the similarity matrix or the mean time series for each of the 111 regions of the Harvard-Oxford atlas \cite{b12}. Importantly, another work, \cite{b19}, based on GCN reported a high variability in the accuracy achieved across sites (50\% to 90\%). Finally, the highest accuracy in classifying participants in the ABIDE database so far was obtained by \cite{b18} by using an ensemble learning strategy on GCN. In this work, the feature vector was represented by the 3D map of the pairwise linear correlations between each atlas’ region and the rest of the brain. A classification accuracy of 73.3\% was achieved by pooling together the results obtained with different atlases featuring different granularity in sampling the cerebral cortex. This was also the only work so far to consider the entire ABIDE-I+II dataset: the final model trained on the ABIDE-I (N=1112) achieved an accuracy of 71.7\% on the ABIDE-II dataset (N=1057).\\
\\
In all of the above work rsfMRI signals were used albeit by summarizing the activity in time to a summary value. We hypothesize that similarity matrices, which only captures first-order statistics between two time-series, will overlook the rich non-linear interactions between them. Therefore in this work we use the original (preprocessed) time courses of spontaneous brain activity as features of interest, and focus on improving the computational tractability of this high-dimensional type of data. As a proof-of-concept, we present an approach which improves on previous work by requiring a very simple neural network architecture which uses minimally preprocessed data and can be trained on NVIDIA Pascal GPU in $<2$ minutes.

\section{ABIDE database and preprocessing}
\label{preprocess}

{\bf The ABIDE I+II} dataset is a collection of structural (T1-weighted) and functional (resting-state fMRI) brain images aggregated across 29 institutions \cite{b16}, available for download\footnote{Download at http://fcon\_1000.projects.nitrc.org/indi/abide/}. It includes 1028 participants with a diagnosis of Autism, Asperger or PDD-NOS (ASD), and 1141 typically developing participants (TD). In people for which the DSM-IV subtyping is available, the ASD sample is composed of 61\% autistic, 25\% Asperger and 14\% PDD-NOS participants. Virtually all the ASD participants are high functioning (99.95\% with IQ \textgreater 70). A large proportion of participants are male adolescents (median age 13 years) although the total age span is between 5 and 64 years of age. Importantly, 20\% of the subjects are females, 1/3 of which with diagnosis of ASD, which represents an important addition with respect to most previous autism studies which focused on the male population exclusively.  The rsfMRI image acquisition time ranges from 2 to 10 minutes, with 85\% of the datasets meeting the suggested duration for obtaining robust rsfMRI estimates \cite{b13}. We chose to cut off the minimum scan duration to 100 time points, which led us to include ~96\% of the whole ABIDE I+II dataset (N=2085, N[ASD]=993), the vast majority of which (95\%) with a minimum acquisition time of 4 minutes.\\
\\
{\bf T1-weighted images} are fed into FAST \cite{b9} for estimation of the bias field and subsequent correction of radiation field (RF) inhomogeneities. This improves the quality of the automated skull stripping performed using bet \cite{b5}.\\
\\
{\bf RSfMRI images:} The first part of the preprocessing pipeline for rsfMRI images is performed using FSL Feat \cite{b8} and consists of standard (1) motion correction using MCFLIRT \cite{b3}, (2) slice timing correction - according to the acquisition parameters specified by the user - (3) spatial smoothing with full width at half maximum (FWHM) equal to 1.5 times the largest voxel size dimension, (4) highpass filtering using a Gaussian-weighted local fit to a straight line. At this time, we also estimate the affine transformation matrix to register the functional data to the MNI152 template in 4x4x4 mm resolution, by combining the transformation matrix of the median functional volume to the skull-stripped T1w, and that of the latter to the MNI152 template provided by FSL. The next steps of rsfMRI preprocessing aim at removing confounding signals from each voxels, which can later yield to artifactual estimation of functional connectivity \cite{b6, b7}. To achieve this, we build a confound matrix containing (1) the 6 estimated motion parameters obtained from the previously performed motion correction, (2) the first eigenvariate of the white matter (WM) and cerebrospinal fluid (CSF), (3) the first derivative of the previous 8 parameters. To estimate the WM and CSF signal, we first register the T1w image to the functional data, and perform a 3-classes tissue segmentation using FAST. The partial volume estimates of WM and CSF are then thresholded to a probability value of 0.9, binarized and used as ROIs to estimate the first eigenvariate of either tissues' time course. The latter step is performed using FSL fslmeants using the --eig option. Finally, the first derivative of the motion parameters, WM and CSF signal is calculated by linear convolution between their time course and a [-1 0 1] vector. Before regressing out the 16-elements confound matrix from the functional data, an additional motion parameter estimation and removal step is performed using ICA AROMA \cite{b4}. Finally, fsl\_glm is used to remove the variance uniquely associated with WM, CSF, and potential residual motion-related signal using the estimated confound matrix. As a last step of the preprocessing pipeline, functional data are bandpass filtered in the range between 0.009 and 0.08 Hz to retain the frequencies which mostly contribute to the part of the resting-state signal which is hypothesized to reflect neuronal interaction \cite{b4}.

\section{Approach}
\label{app}
As alluded to in \ref{intro}, one of the main problem with neuroimaging data is its dimensionality. In order to apply deep learning to this high-dimensional data we have to reduce this dimensionality. Since we are dealing with rsfMRI data, we can either summarize the entire time-series per voxel into a single number or we can summarize the activities of several contiguous voxels. An example of the former is the widely used approach of calculating the correlation between the rsfMRI signal between any two voxels. As previously mentioned, this approach assumes that only fist-order statistics between time-series represents relevant features, and discards potentially informative non-linear interaction terms\cite{b21}. In this study we therefore chose to maintain the full time course as source of features, and to aggregate it across meaningful ensemble of voxels.\\
\\
In order to summarize a group of ``meaningful'' voxels together we utilized several anatomical and functionally derived atlases, which divide the brain in sets of voxels featuring relatively similar activity. We used (i) the Automated Anatomical Labeling (AAL, or anatomical automatic labeling) atlas based on the anatomical parcellation of the a single subject in a standard space \cite{b10}, (ii) the Schaefer-100 and Schaefer-400, created using rsfMRI data from 1489 subjects co-registered using surface-based alignment, where a gradient weighted Markov random field approach was employed to identify 100 and 400 parcels, respectively\cite{b10}, (iii) the Harvard-Oxford probabilistic \cite{b12} atlas, covering 48 cortical and 21 subcortical areas, manually-delineated on T1-weighted brain images.\\
\\
Our approach is therefore as follows:

\begin{itemize}
    \item preprocess the rsfMRI as in \ref{preprocess}
    \item for a given atlas (AAL, Schafer-100 etc.,) extract the mean time-series within each ROI
    \item the matrix of these time-series are input into a 1-D convolutional network with each channel corresponding to a different ROI.
    \item binary classification was performed between controls and patients, and the error backpropagated.
\end{itemize}

\section{Experiments \& Results}
\label{expnres}
\subsection{Network Architecture}
The network consist of a 1-D convolution layer with number of channels equal to the number of ROIs depending on the atlas used. Following a single convolution layer was the adaptive average pooling layer that averaged the values across an entire channel. Finally, the output is flattened and fed to  a fully connected layer, whose output was then transformed by a softmax function to generate the probabilities of each class.  A dropout factor of 0.2 was applied to the fully connected layer to regularize the network.
The network is trained using Adam optimizer with a learning rate of 0.0001 and a weight decay of 0.002.  The data is split into training (70\%), validation (10\%) and testing (20\%), and the model is selected based upon the best validation loss.

\subsection{Results}
ABIDE I+II dataset is collected from different sites at different scanning time and different TR.
To ensure the consistency of the signals we cut-off the time series signals at different number of time points thresholds and evaluate the performance of the model at each cut-off threshold for different atlas parcelations. Subjects with number of time points less than cut-off threshold were discarded.\\

\begin{figure}[htbp]
\centerline{\includegraphics[width=0.5 \textwidth]{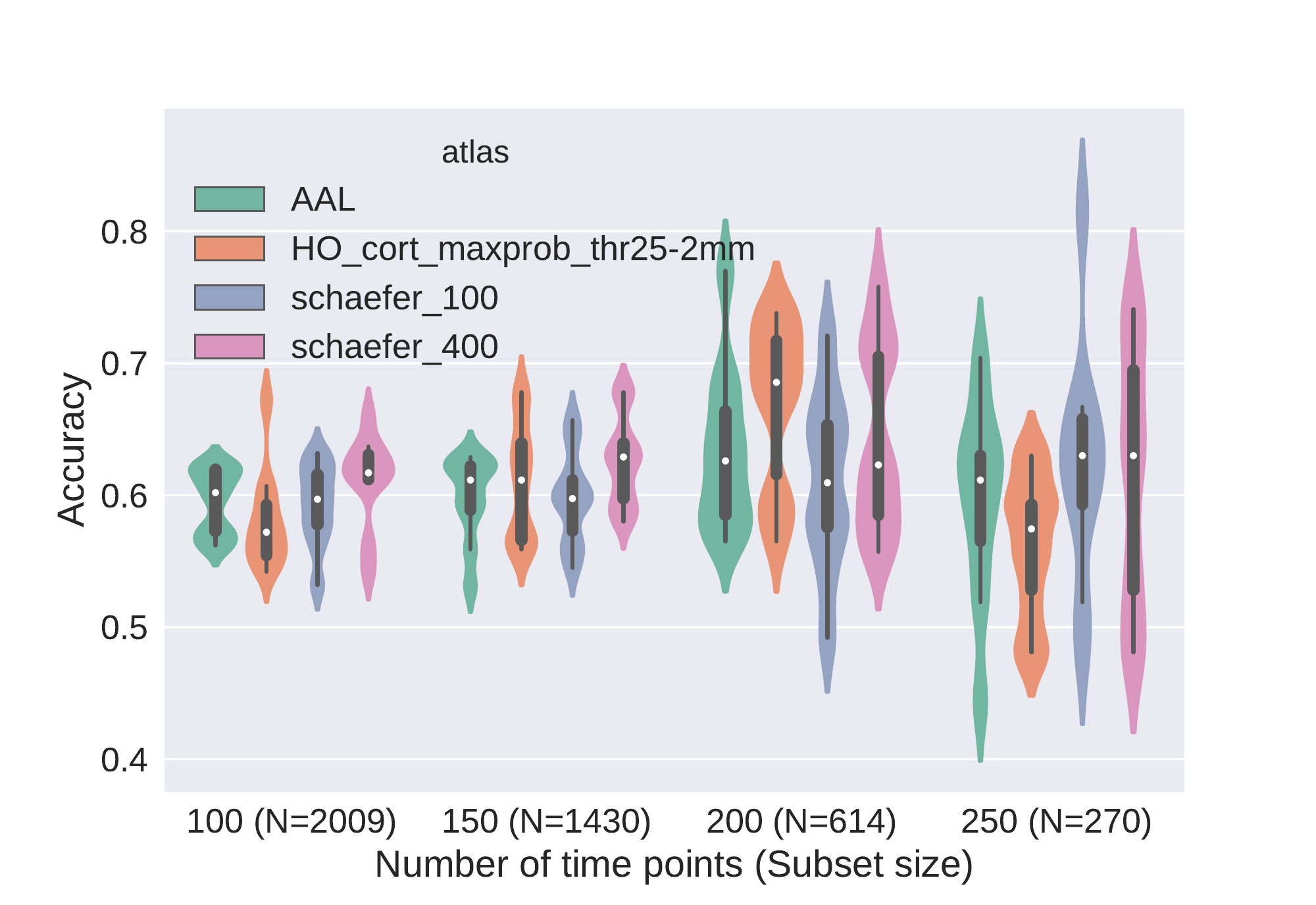}}
\caption{10-fold cross validation accuracies using AAL, Harvard Oxford, Schaefer-100 and Schaefer-400 atlases for training and testing at different number of time courses time points on ABIDE I+II.
The results illustrates the superiority of using Harvard Oxford atlas at 200 time points. }
\label{graph}
\end{figure}

\paragraph{10-fold cross validation} We used a 10-fold cross validation scheme to report the accuracies in Fig.~\ref{graph}. We observed a positive correlation between the number of time points used and the performance of the model across all atlases. The accuracy of the model degrades after the 200 time points threshold, which is presumably due to the limited number of samples that remains for training of the model. Another observation is the superior performance of time series extracted from Harvard Oxford at 200 time points against other atlases with a cross validated test accuracy of 68\%.

\begin{figure}[htbp]
\centerline{\includegraphics[width=0.5\textwidth]{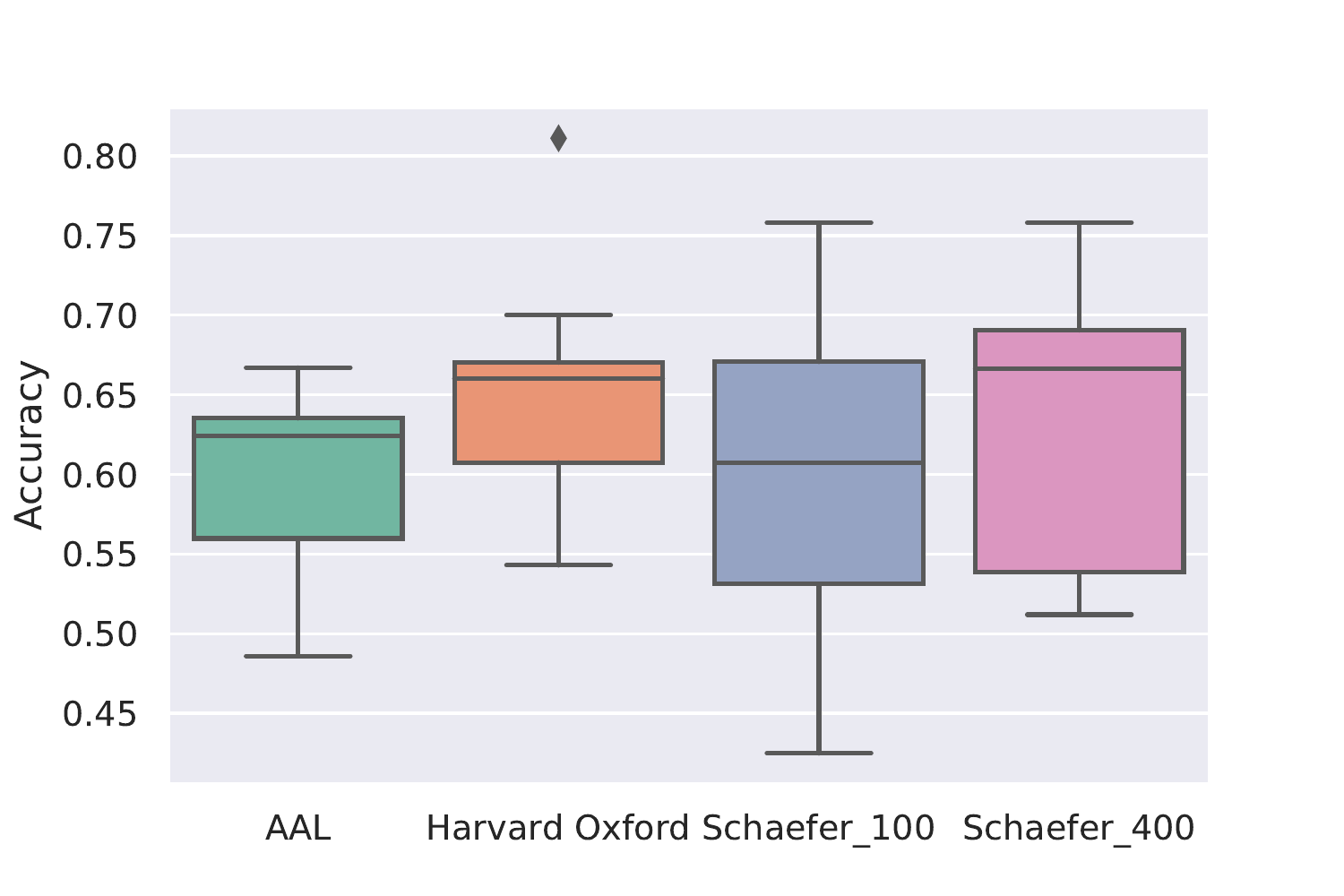}}
\caption{Leave site out validation results using AAL, Harvard Oxford, Schaefer-100 and Schaefer-400 atlases at 200 time points. We observe that Harvard Oxford provides the highest average test accuracy across sites and lowest variance.}
\label{site_out}
\end{figure}

\paragraph{Across sites cross validation} To evaluate classifier performance across sites, we performed a leave-one-site-out cross validation scheme using the four atlases atlas at a cut-off of 200 time points threshold. This process excluded data from one site from the training process, and used that data as the test set to evaluate the model. The rationale was to test the applicability of our trained model for new, different sites. The results of these further analyses are reported in Fig. \ref{site_out}. The results showed that Schaefer-400 and Harvard Oxford atlases yielded the highest mean classification accuracy across sites with 64.3\% and 65.1\% respectively. However, using the Harvard Oxford atlas for classification demonstrated lower variance across sites which entitles it to be more applicable in practical environments regardless of site of acquisition. We thus selected the Harvard Oxford atlas at 200 timepoints to undergo our further analysis and report the details of the leave-one-site validation results in Table \ref{tab1}.

\begin{table}[b]
\caption{Leave-site-out cross validation}
\begin{center}
\begin{tabular}{|c|c|c|c|c|}
\hline
%\textbf{Site-out}&\multicolumn{3}{|c|}{\textbf{Table Column Head}} \\
\cline{2-4} 
\textbf{Site-out} & \textbf{\textit{N (ASD/TD)}} & \textbf{\textit{Acc.}}& \textbf{\textit{Sens.}}& \textbf{\textit{Spec.}} \\
\hline

ABIDEII-ETH1 & 37 (24/13) &0.81 &0.77 &0.83\\
OLIN & 36 (16/20) &0.72 &0.65 &0.81\\
LEUVEN2 & 35 (20/15) &0.69 &0.8 &0.6\\
USM & 101 (43/58) &0.69 &0.66 &0.74\\
CMU & 27 (13/14) &0.67 &0.57 &0.77\\
UM1 &  110 (55/55) &0.66 &0.76 &0.55\\
UM2 &  35 (22/14) &0.66 &0.69 &0.64\\
STANFORD & 22 (20/2) &0.64 &1 &0.6\\
ABIDEII-USM1 & 33 (16/17) &0.64 &0.76 &0.5\\
ABIDEII-TCD1 & 41 (21/20)  &0.63 &0.5 &0.76\\
ABIDEII-OILH2 & 59 (35/24)  &0.63 &0.54 &0.69\\
ABIDEII-IU1 & 40 (20/20)  &0.6 &0.6 &0.6\\

\hline
%\multicolumn{4}{l}{$^{\mathrm{a}}$Sample of a Table footnote.}
\end{tabular}
\label{tab1}
\end{center}
\end{table}

\subsection{Dataset  Heterogeneity}
ABIDE I+II is a multi-site dataset with different scanning parameters and different population characteristics. To address the effect of this heterogeneity on our results, we assessed the influence of site and sample variability by post-hoc analysis.\\

\begin{figure}[htbp]
\centerline{\includegraphics[width=0.5\textwidth]{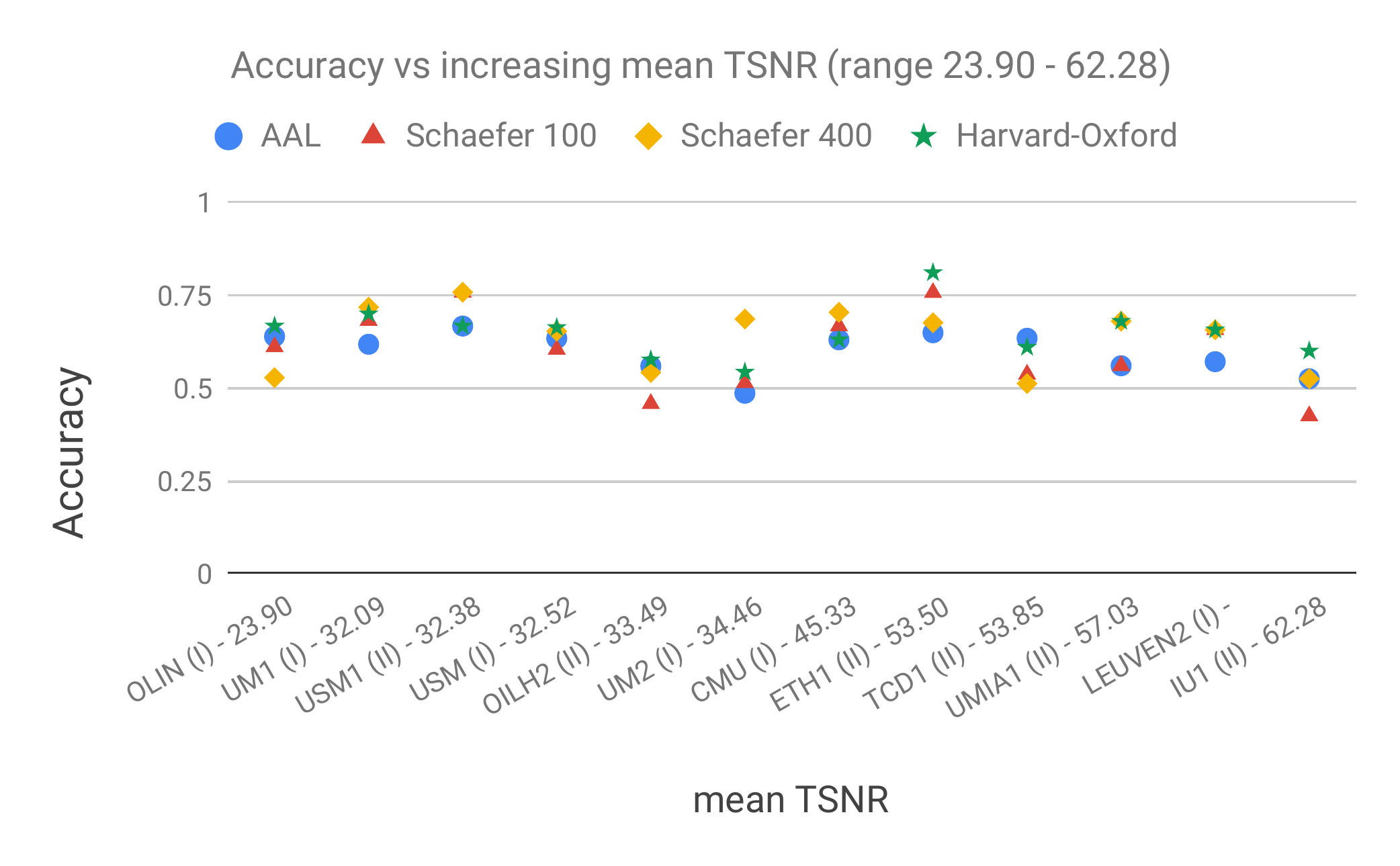}}
\caption{Increasing mean TSNR across sites vs leave site out validation accuracy using different atlases at 200 time points. The results shows no association between classifcation accuracy and mean TSNR values ($p>0.41$ for all atlases).}
\label{accuracys_vs_tsnr}
\end{figure}

\paragraph{Site variability}
ABIDE I+II features 29 different sites, each using different scanning hardware and sequences, hence resulting in different image quality and resolution. 
The latter does not represent a concern, as (1) all images were resampled to the same 4x4x4 mm voxel size in MNI space; (2) the atlas seed regions from which the mean time courses were extracted are at least 10 times bigger that the size of the original voxel; (3) potential partial volume contamination are accounted for by WM and CSF signal regression. To quantify differences in image quality, we estimated temporal signal-to-noise ratio (TSNR) of the raw 4D rs-fMRI images, and plotted it against validation accuracies in the leave-site out validation among different atlases at 200 time points in Fig \ref{accuracys_vs_tsnr}. Despite substantial variability in TSNR across sites, we did not observe a significant association between classification accuracy and image quality (Pearson correlation coefficient: $p > 0.41$ for all atlases).

As TSNR varied considerably also within sites, we further assessed whether this variable would have an impact on the proportions reported in the test matrix. This analysis focused on the k-fold evaluation of HO atlas at 200 time points. A one-way ANOVA on the mean TSNR  revealed no significant differences for images attributed to different elements of the confusion matrix ($F_{610,3} = 0.623, p > 0.59$). In particular, this shows that on average false positives/negatives did not feature a lower TSNR with respect to correctly classified ASD/TD people.\\

\paragraph{Effect of Age and Gender}
In autism, symptom expression and severity is related to age and gender. Therefore we performed additional analyses to assess the impact of these two variables on the results of classification.

Autism is largely a male condition. This is reflected also in the ABIDE dataset: the subset of participants we considered consists of 1754 males ($N_{ASD}=865, N_{TD}=889$) and 415 females ($N_{ASD}=276, N_{TD}=139$). However, when we assessed the effect of gender on the class attribution carried out by our network, we did not detect any significant gender difference in true/false positives/negatives ($\chi^2_{(3,1)} = 5.74, p>0.12$). 

Regarding age differences, while the dataset spans from 6 to 64 years old, about 76\% subjects are below the age of 20 years old. This makes age binning difficult, therefore we trained our network using the entire age distribution and again carried out post-hoc analysis on the confusion matrix, to assess whether the classification could have been influenced by Age. A one-way ANOVA revealed no significant differences in the mean age of the participants across the four elements of the confusion matrix ($F_{610,3} = 0.071, p > 0.96$).

In summary, these post-hoc analysis suggest variability in age and gender did not have a large influence on the classification carried out by our neural network.

\section{Conclusion}
\label{con}
We have proposed an elegant and simple machine learning solution for using rsfMRI using preprocessing and transformations that do not compromise its temporal properties.\\
\\
Resting-state fMRI data is often dealt with in machine learning (or deep learning) by summarizing the temporal activity as a correlation (or its derivatives). Correlation is a first-order transformation, which does not account for higher order interactions between time courses. In order to retain all the temporal information and yet deal with the high dimensionality of the data, we summarized a group of ``meaningful" voxels leading to a time series per region-of-interest.\\
\\
In this proof-of-concept paper we were able to show that for a challenging (heterogeneous) data set like the ABIDE, we were able to achieve cross-validated accuracies of 68\%. There was only a slight reduction in average accuracy to 65.1\% when testing the model on data from new sites using leave-one-site out cross-validation. The whole pipeline required minimal preprocessing and the network was trained on NVIDIA Pascal GPU in less that 2 minutes.\\
\\
In the future, we will systematically study this approach and apply it to other psychiatric disorders. A systematic effect of the various atlases and the duration of the resting-state time-series will be investigated.  \\

\section*{Acknowledgement}
This work was supported by the Netherlands Organization for Scientific Research (NWO/ZonMW Vidi 016.156.318).

\end{document}